\documentclass{aa}  
\usepackage{color}
\usepackage{graphicx}
\usepackage{lscape}
\usepackage{natbib}
\usepackage[varg]{txfonts}
\usepackage{url}
\usepackage{xspace}
\bibpunct{(}{)}{;}{a}{}{,}
\def\etal{{et\,al.}\ }
\newcommand{\Teff}{$T\mathrm{\hspace*{-0.4ex}_{eff}}$\,}
\newcommand{\logg}{$\log\,g$\hspace*{0.5ex}}

\def\hd{HD\,149499\,B}
\def\pgnull{PG\,0109+111}
\def\re{RE\,0503$-$289}
\begin{document}

\title{First detection of bromine and antimony in hot stars
  \thanks{Based on observations made with the NASA-CNES-CSA Far
    Ultraviolet Spectroscopic Explorer.} \thanks {Based on
    observations with the NASA/ESA Hubble Space Telescope, obtained at
    the Space Telescope Science  Institute, which is operated by the
    Association of Universities for Research in Astronomy, Inc., under
    NASA contract NAS5-26666.}}

\author{K\@. Werner\inst{1} \and T\@. Rauch\inst{1} \and M\@. Kn\"orzer\inst{1} \and J.W\@. Kruk\inst{2}}

\institute{Institute for Astronomy and Astrophysics, Kepler Center for Astro and
Particle Physics, Eberhard Karls University, Sand~1, 72076
T\"ubingen, Germany\\ \email{werner@astro.uni-tuebingen.de}
\and
           NASA Goddard Space Flight Center, Greenbelt, MD\,20771, USA
}

\date{Received 29 January 2018 / Accepted 5 March 2018}

\authorrunning{K. Werner \etal}
%\titlerunning{}

\abstract{Bromine (atomic number $Z=35$) and antimony ($Z=51$) are
  extremely difficult to detect in stars. In very few instances, weak
  and mostly uncertain identifications of \ion{Br}{i}, \ion{Br}{ii},
  and \ion{Sb}{ii} in relatively cool, chemically peculiar stars were
  successful. Adopted solar abundance values rely on meteoritic
  determinations. Here, we announce the first identification of these
  species in far-ultraviolet spectra of hot stars (with effective
  temperatures of 49\,500--70\,000\,K), namely in helium-rich (spectral
  type DO) white dwarfs. We identify the \ion{Br}{vi} resonance line
  at 945.96\,\AA. A previous claim of Br detection based on this line
  is incorrect because its wavelength position is inaccurate by about
  7\,\AA\ in atomic databases. Taking advantage of precise laboratory
  measurements, we identify this line as well as two other,
  subordinate \ion{Br}{vi} lines. Antimony is detected by the
  \ion{Sb}{v} resonance doublet at 1104.23/1225.98\,\AA\, as well as
  two subordinate \ion{Sb}{vi} lines.  A model-atmosphere analysis
  reveals strongly oversolar Br and Sb abundances that are caused by
  radiative-levitation dominated atomic diffusion.}


\maketitle
%
%________________________________________________________________

\section{Introduction}
\label{intro}

Bromine (atomic number $Z=35$) and antimony ($Z=51$) are rather rare
elements in the Universe and hard to detect in stars. Even the adopted
solar abundance values \citep[number ratios Br/H = $3.5 \times
  10^{-10}$, Sb/H = $1.0 \times 10^{-11}$;][]{2009ARA&A..47..481A}
were established indirectly from meteoritic measurements. Using
improved analysis methods, it has been shown recently that the heavy halogen
abundances in chondritic meteorites are significantly lower than
previously thought and, for bromine in particular, this amounts to a
factor of nine \citep{2017Natur.551..614C}.

The detection of bromine in stars succeeded only recently.
\cite{2004A&A...425..263C} and \cite{2006A&A...447..681C} have
identified \ion{Br}{ii} lines in the optical spectra of the
mercury-manganese (HgMn) star HR\,7143 and the He-weak chemically
peculiar (CP) star 3\,Cen\,A, respectively, and bromine excesses of
2.3 and 2.6\,dex were measured. \ion{Br}{i} lines were reported in the
spectrum of the very peculiar star HD\,101065 (alias Przybylski's
star) by \citet{2005ASPC..336..309B}. There is hardly any detection of
antimony in stars, and respective claims are considered uncertain.
Weak \ion{Sb}{ii} lines were reported in ultraviolet (UV) spectra of
the HgMn star $\chi$\,Lupi \citep[][concluding an Sb excess of
  1.6\,dex]{1999AJ....117.1454L} and in the hot-Am star HR\,3383
\citep{2008CoSka..38..463W}. In hotter stars, at higher ionization
stages, bromine and antimony were not detected up to now. Here, we
announce the identification of \ion{Br}{vi}, \ion{Sb}{v}, and
\ion{Sb}{vi} lines in three hot white dwarf stars.

\begin{figure}[t]
 \centering
 \includegraphics[width=0.9\columnwidth]{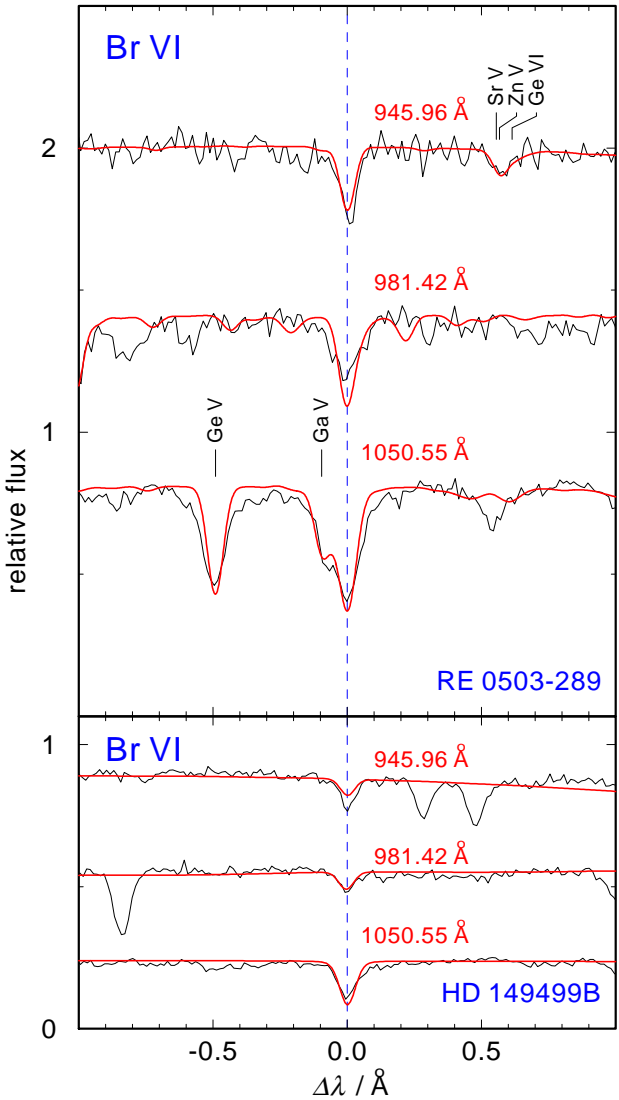}
  \caption{Lines of \ion{Br}{vi} in the DO white dwarfs \re\ (\emph{top}
    panel) and \hd\ (\emph{bottom}).}\label{fig:br}
\end{figure}

\begin{figure}[t]
 \centering \includegraphics[width=0.9\columnwidth]{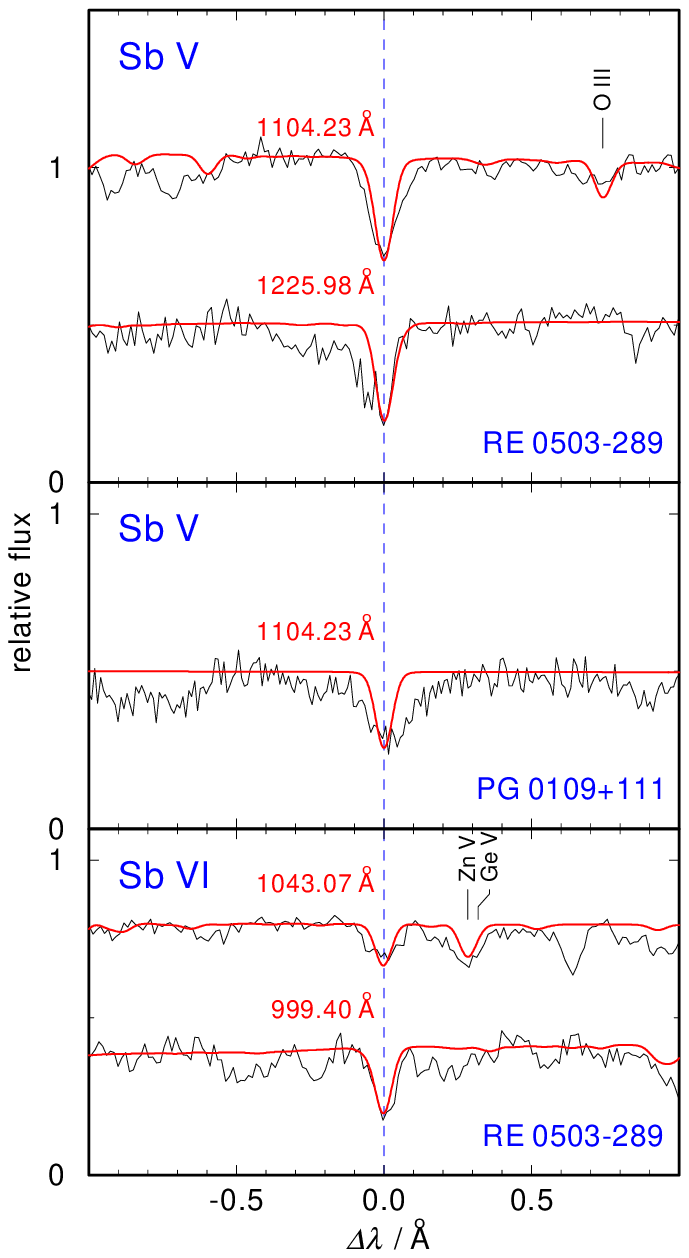}
  \caption{Lines of \ion{Sb}{v} in the DO white dwarf \re\ (\emph{top} panel)
    and \pgnull\ (\emph{middle}), as well as two \ion{Sb}{vi} lines in
    \re\ (\emph{bottom}).}\label{fig:sb}
\end{figure}

\begin{table*}[t]
\begin{center}
\caption{Spectral lines of Br and Sb detected in three He-rich white dwarfs\tablefootmark{a}. }
\label{tab:lines} 
\small
\begin{tabular}{lccrrrccc}
\hline 
\hline 
\noalign{\smallskip}
Ion          & Transition                           & $J-J$     & $\lambda_{\rm lab}$/\AA & $\log g_if_{ik}$ & $E_i$/cm$^{-1}$ & \re    & \pgnull & \hd \\ 
\hline 
\noalign{\smallskip}
\ion{Br}{vi} & 4s$^2$\,$^1$S -- 4s4p\,$^3$P$^{\rm o}$ & $0-1$     & 945.96                &  $-2.193$       & 0.0            &$\times$&         &$\times$ \\
\             & 4s4p $^1$P$^{\rm o}$ -- 4p$^2$\,$^3$P  & $1-2$     & 981.42                &  $-0.600$       & 151\,285.0     &$\times$&         &$\times$ \\
\             &                                      & $1-1$     & 1050.55               &  $-0.275$       & 151\,285.0     &$\times$&         &$\times$\\
\noalign{\smallskip}
\hline
\noalign{\smallskip}
\ion{Sb}{v}  & 5s\,$^2$S -- 5p\,$^2$P$^{\rm o}$       & $1/2-3/2$ & 1104.20               &  $0.052$        & 0.0            &$\times$&$\times$\\
\             &                                      & $1/2-1/2$ & 1225.98               &  $-0.232$       & 0.0            &$\times$& \\
\ion{Sb}{vi} & 5s\,$^3$D -- 5p\,$^3$F$^{\rm o}$       & $3-4$     & 999.40                &  $0.440$        & 242\,916.8     &$\times$& \\
\             & 5s\,$^3$D -- 5p\,$^3$P$^{\rm o}$       & $1-0$     & 1043.07               &  $0.520$        & 242\,916.8     &$\times$&\\ 
\noalign{\smallskip} \hline
\end{tabular} 
\tablefoot{\tablefoottext{a}{Excitation energies $E_i$ for \ion{Br}{vi} from
\citet{2012JQSRT.113.2072R}, for \ion{Sb}{v} from
\citet{1966PhDT........53C}, and for \ion{Sb}{vi} from
\citet{2000PhyS...61..420C}. Oscillator strengths $f_{ik}$ for
\ion{Br}{vi} from \citet{2012JQSRT.113.2072R}, for \ion{Sb}{v} from
\citet{2000ApJS..130..403M}. For \ion{Sb}{vi}, we used the $f_{ik}$
values of the respective lines in the isoelectronic \ion{Ge}{v} ion
from \citet{test}. $g_i$ is the statistical weight of lower level
$i$. An entry ``$\times$'' in the last three columns indicates which
line was detected in which star.}} 
\end{center}
\end{table*}

\section{Line identification and atomic data}
\label{sect:atoms}

We investigated UV spectra of three hot helium-rich white dwarfs
(spectral type DO). In all of them, heavy elements beyond the iron
group (atomic number $Z>29$) were detected previously. In
\hd\ \citep[effective temperature \Teff = 49\,000\,K, surface gravity
  $\log$ ($g$/cm\,s$^{-2}$) = 7.97; ][]{1995A&A...300L...5N},
\citet{2005ApJ...630L.169C} have identified six such species (Ge, As,
Se, Sn, Te, and I). Eight trans-iron elements (Zn, Ga, Ge, Se, Sr, Sn,
Te, I) were found in \pgnull\ \citep[\Teff = 70\,000\,K, \logg =
  8.0;][]{2018arXiv180102414H}. The third DO discussed here
\citep[\re, \Teff = 70\,000\,K, \logg = 7.5; ][]{1996A&A...314..217D}
is a truly outstanding object with respect to its heavy-element
variety. Fourteen trans-iron elements (see below) were identified and
abundances determined \citep[][and references
  therein]{2017A&A...606A.105R}. Generally, large or extreme
overabundances up to five dex oversolar were found, most probably
caused by radiative levitation \citep{rauchetal2016mo}.

For our assessment of \re\ and \pgnull, we used spectra taken with
the Far Ultraviolet Spectroscopic Explorer (FUSE) and the
\emph{Hubble} Space Telescope (HST). For details of the observations,
we refer the reader to our earlier work \citep{2017A&A...606A.105R,2018arXiv180102414H}. In addition, we used co-added
archival FUSE spectra of \hd. In total, three Br and four Sb lines
were identified (Table~\ref{tab:lines}, Figs.\,\ref{fig:br} and
\ref{fig:sb}). 

\subsection{Bromine}

The first laboratory investigation of the \ion{Br}{vi} spectrum in the
far-UV wavelength region was performed by
\citet{1934PPS....46..163R}. In particular, the wavelength of the
4s$^2$\,$^1$S$_0$ -- 4s4p $^3$P$^{\rm o}_1$ resonance
(intercombination) line was determined to 939.57\,\AA. The energy
levels derived from these early measurements are currently still
listed in the NIST database \citep[but classified as ``not critically
  evaluated by NIST'',][]{NIST_ASD} and were used, for example, in the
linelists established by \citet{1987JPCRD..16S....K} and
\citet{2000ApJS..130..403M} and, thus, are widely used in atomic
databases. It had, however, been ignored that, not earlier than half a
century after the first investigation, \citet{1984PhLA..105..212C}
made a new measurement of that line and found the wavelength
6.7\,\AA\ longer (946.3$\pm$0.2\,\AA), suggesting that the excitation energies of all triplet levels in \ion{Br}{vi} by
756\,cm$^{-1}$ should be  adjusted downward. \citet{1986PhyS...34..135J} further improved the line
position measurements of \ion{Br}{vi} and found 945.966\,\AA\ for the
resonance line. In a more recent work, new line spectra measurements
were performed and a revised and extended analysis of \ion{Br}{vi} was
published by \citet{2012JQSRT.113.2072R}, with a wavelength of
945.960$\pm$0.005\,\AA. As a consequence of this inaccuracy in the
databases, \citet{2005ApJ...630L.169C} have erroneously identified a
line in the spectrum of the cool DO \hd\ located at 939.6\,\AA\ as due
to this \ion{Br}{vi} line. We have identified it in the spectra of \re\  and
\hd\ at the correct wavelength (Fig.\,\ref{fig:br}). We cannot offer
an identification for the line at 939.6\,\AA.

Two more \ion{Br}{vi} lines can be identified in \re\ and (weaker) in
\hd. They are the two strongest components ($J=1-1$ and $1-2$) of the
4s4p $^1$P$^{\rm o}$ -- 4p$^2$\,$^3$P multiplet at 1050.55 and
981.42\,\AA\ (Fig.\,\ref{fig:br}).

\subsection{Antimony}

According to the NIST database and the compilation of
\citet{2000ApJS..130..403M}, the \ion{Sb}{v} 5s\,$^2$S --
5p\,$^2$P$^{\rm o}$ resonance doublet is located at 1104.32 and
1226.00\,\AA. These values are from level energies (also given in
NIST) that were established by \citet{1927PNAS...13..341L} and
\citet{1931PPS....43..538B} based on laboratory work of
\citet{1929PhRv...34..406G} \citep[see][]{1971stas.book.....M}. For his
PhD thesis, \cite{1966PhDT........53C} performed a very detailed
analysis of antimony spectra, which is disregarded in the compilations
of \citet{1971stas.book.....M} and NIST. He derived a more accurate
wavelength of 1104.235\,\AA\ for the blue resonance component. This
deviates by just 0.04\,\AA\ from the line position in our WD spectrum,
which is within the wavelength uncertainty of the FUSE
observation. For the red component, \cite{1966PhDT........53C} gives
1225.983\,\AA, very close to the original value of 1226.00\,\AA. The
doublet is identified in \re\ and the blue component in \pgnull\ (the
observations do not cover the red component), see Fig.\,\ref{fig:sb}.

Two other \ion{Sb}{v} lines at 1506/1524\,\AA\ are listed in the NIST
database. They were originally thought to be the 5d--5f doublet. They are, however, misidentifications in the early work of
\citet{1927PNAS...13..341L}, as was already pointed out by
\citet{1931PPS....43..538B}. They are no \ion{Sb}{v} lines.

We also identify a prominent \ion{Sb}{vi} line in \re\ at
944.40\,\AA. According to the level energies in the NIST database
\citep[which are from][]{2000PhyS...61..420C}, it is the 5s\,$^3$D$_3$
-- 5p\,$^3$F$^{\rm o}_4$ transition. This corresponds to the
isoelectronic \ion{Sn}{v} line at 1160.74\,\AA, which is of similar
strength in \re\ \citep{2012ApJ...753L...7W}. The other components of
this \ion{Sb}{vi} multiplet are expected to be weaker and,
occasionally, they are blended with other photospheric lines. Instead,
we identify the $J=1-0$ component of the 5s\,$^3$D -- 5p\,$^3$P$^{\rm
  o}$ multiplet at 1043.07\,\AA\ (Fig.\,\ref{fig:sb}).

\section{Spectral analysis}

We performed a quantitative spectral analysis to derive the abundances
of Br and Sb from detailed line-profile fits. To this end, we used the
T\"ubingen Model-Atmosphere Package
(TMAP\footnote{\url{http://astro.uni-tuebingen.de/~TMAP}}) to compute
non-local thermodynamic equilibrium (NLTE), plane-parallel,
line-blanketed atmosphere models in radiative and hydrostatic
equilibrium
\citep{1999JCoAM.109...65W,2003ASPC..288...31W,tmap2012}. For \hd, we
computed models including H, He, and Br. In a final formal solution of
the radiation transfer equation, line profiles were calculated
accounting for fine-structure splitting. The same procedure was
performed for \re\ and \pgnull, but in these cases we used our
detailed metal-line blanketed model atmospheres from previous
work. The \re\ model includes 26 species \citep{2017A&A...606A.105R}
plus Br and Sb, and the \pgnull\ model four species \citep[He, C, N,
  and O;][]{2018arXiv180102414H} plus Sb. 

Our model atoms for Br and Sb consist of ionization stages
IV--VII. The numbers of NLTE levels/lines per ion are 15/1, 5/4,
20/42, and 1/0 for \ion{Br}{iv}--\ion{Br}{vii}, respectively. Level
energies and oscillator strengths $f_{ik}$ for \ion{Br}{v} and
\ion{Br}{vi} were taken from \citet{2014JQSRT.147...86R} and
\citet{2012JQSRT.113.2072R}, respectively. Level energies for
\ion{Br}{iv} were adopted from NIST and the $f_{ik}$ value for the
considered resonance line is from \citet{1969MNRAS.142..265W}. 

For Sb, the numbers of NLTE levels/lines per ion are 3/1, 5/1, 8/1, and
1/0 for \ion{Sb}{iv}--\ion{Sb}{vii}, respectively. Level energies for
\ion{Sb}{iv} were taken from NIST, for \ion{Sb}{v} from
\citet{1966PhDT........53C}, and  for \ion{Sb}{vi} from
\citet{2000PhyS...61..420C}. The $f_{ik}$ values of the considered
\ion{Sb}{iv} and \ion{Sb}{v} resonance lines are from
\citet{2000ApJS..130..403M}. That of the considered \ion{Sb}{vi}
resonance line at 285\,\AA\ is from \citet{2000PhyS...61..420C}. The
two observed UV lines are subordinate and they were considered in the
final spectrum synthesis caluclation, only. For their $f_{ik}$ values
see Table~\ref{tab:lines}.

For both species, Br and Sb, bound-free cross sections were assumed
to be hydrogen-like. Line profiles for quadratic Stark broadening and
electron collisional rates were computed with usual approximate
formulae \citep[see, e.g.,][]{2012ApJ...753L...7W}.

\begin{figure}[t]
 \centering \includegraphics[width=\columnwidth]{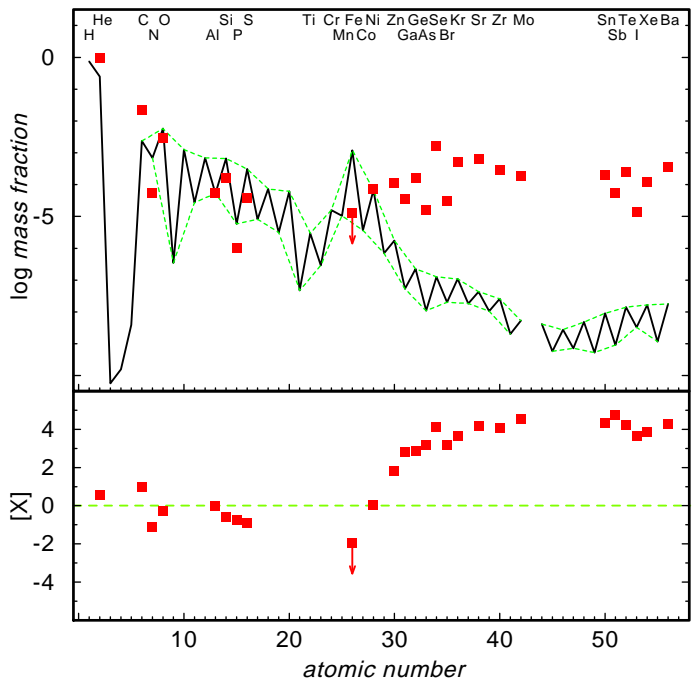}
  \caption{Elemental abundances in the hot white dwarf \re. \emph{Top}:
   Black line indicates solar abundances. \emph{Bottom}:
    Logarithm of mass fractions relative to solar values. The green dashed
    line indicates solar abundances. All results except for Br and Sb
are    taken from \citet{2017A&A...606A.105R} and references
      therein. For iron, an upper limit is indicated by the arrow.}\label{fig:abu}
\end{figure}

For each star, several model atmospheres with different Br and Sb
abundances were computed. Our best fits to the observed line profiles
are displayed in Figs.\,\ref{fig:br} and \ref{fig:sb}. The resulting
element abundances are summarized in Table~\ref{tab:results}. Analysis
errors are estimated to $\pm0.3$\,dex, but a similar systematic error
for Sb must be accounted for because of the atomic data approximations
in our model atom. For \re, the results are displayed in
Fig.\,\ref{fig:abu} together with all element abundances hitherto
measured in this star.

\begin{table}[t]
\begin{center}
\caption{Bromine and antimony abundances measured in He-rich white
  dwarfs\tablefootmark{a}. }
\label{tab:results} 
\small
\begin{tabular}{ccccc}
\hline 
\hline 
\noalign{\smallskip}
Star    & \Teff/K & \logg  & Br   & Sb \\ 
\hline 
\noalign{\smallskip}
\re     & 70\,000  & 7.50   & $-4.5$ & $-4.3$ \\
\pgnull & 70\,000  & 8.00   &        & $-5.3$ \\
\hd     & 49\,500  & 7.97   & $-4.3$ &   \\     
\noalign{\smallskip} \hline
\end{tabular} 
\tablefoot{\tablefoottext{a}{Abundances given as logarithm of mass
    fraction. Solar mass fractions are $\log$\,Br$ =-7.7$
    and $\log$\,Sb $=-9.0$. Surface gravity $g$ in cm/s$^2$.}} 
\end{center}
\end{table}

\section{Summary and conclusions}

We reported the first detection of the bromine and antimony in hot
stars. We identified spectral lines from \ion{Br}{vi} as well as
\ion{Sb}{v} and \ion{Sb}{vi} in helium-rich white dwarfs. As a result
of our NLTE model-atmosphere analysis, we found that the Br abundance
in \re\ and \hd\ is 1600 and 2500 times solar and the Sb abundance in
\re\ and \pgnull\ is even 50\,000 and 5000 times solar. 

In the case of \re, Br and Sb are two more trans-iron species in
addition to the fourteen already found. Among all white dwarfs, this
rich diversity of detected heavy elements is outstanding. The Br and
Sb abundances fit into the general trend that all trans-iron elements
are extremely overabundant (Fig.\,\ref{fig:abu}). As we have discussed
earlier, the origin for this phenomenon is most likely rather
efficient radiative levitation \citep[][and references
  therein]{2017A&A...606A.105R}.

As we have demonstrated, our ongoing work on the detection of
trans-iron elements in hot white dwarfs and respective
model-atmosphere analyses repeatedly faces the problem of inaccurate
or lacking atomic data, primarily level energies of moderately
ionized atoms (ionization stages IV to VIII) and oscillator
strengths. For many species, laboratory spectra and their extended
analysis are badly needed. A large number of spectral lines remain
unidentified in hot white dwarfs and it is a reasonable suspicion that
many of these stem from hitherto undetected elements. Good candidates
are species with atomic numbers in the range $Z=43-49$ (Tc--In; as
indicated by Fig.\,\ref{fig:abu}), or even species beyond the most
heavy element discovered so far in any white dwarf ($Z=56$, barium).

\begin{acknowledgements} 
The TMAD tool (\url{http://astro.uni-tuebingen.de/~TMAD}) used for
this paper was constructed as part of the activities of the German
Astrophysical Virtual Observatory. Some of the data presented in this
paper were obtained from the Mikulski Archive for Space Telescopes
(MAST). STScI is operated by the Association of Universities for
Research in Astronomy, Inc., under NASA contract NAS5-26555. Support
for MAST for non-HST data is provided by the NASA Office of Space
Science via grant NNX09AF08G and by other grants and contracts. This
research has made use of NASA's Astrophysics Data System and the
SIMBAD database, operated at CDS, Strasbourg, France.
\end{acknowledgements}

\bibliographystyle{aa}
\bibliography{aa}

\begin{thebibliography}{33}
\expandafter\ifx\csname natexlab\endcsname\relax\def\natexlab#1{#1}\fi

\bibitem[{{Asplund} {et~al.}(2009){Asplund}, {Grevesse}, {Sauval}, \&
  {Scott}}]{2009ARA&A..47..481A}
{Asplund}, M., {Grevesse}, N., {Sauval}, A.~J., \& {Scott}, P. 2009, \araa, 47,
  481

\bibitem[{{Badami}(1931)}]{1931PPS....43..538B}
{Badami}, J.~S. 1931, Proceedings of the Physical Society, 43, 538

\bibitem[{{Bidelman}(2005)}]{2005ASPC..336..309B}
{Bidelman}, W.~P. 2005, in Astronomical Society of the Pacific Conference
  Series, Vol. 336, Cosmic Abundances as Records of Stellar Evolution and
  Nucleosynthesis, ed. T.~G. {Barnes}, III \& F.~N. {Bash}, 309

\bibitem[{{Castelli} \& {Hubrig}(2004)}]{2004A&A...425..263C}
{Castelli}, F. \& {Hubrig}, S. 2004, \aap, 425, 263

\bibitem[{{Chan}(1966)}]{1966PhDT........53C}
{Chan}, C. 1966, PhD thesis, The University of British Columbia (Canada).

\bibitem[{{Chayer} {et~al.}(2005){Chayer}, {Vennes}, {Dupuis}, \&
  {Kruk}}]{2005ApJ...630L.169C}
{Chayer}, P., {Vennes}, S., {Dupuis}, J., \& {Kruk}, J.~W. 2005, \apjl, 630,
  L169

\bibitem[{{Churilov} {et~al.}(2000){Churilov}, {Azarov}, {Ryabtsev},
  {Tchang-Brillet}, \& {Wyart}}]{2000PhyS...61..420C}
{Churilov}, S.~S., {Azarov}, V.~I., {Ryabtsev}, A.~N., {Tchang-Brillet},
  W.-{\"U}.~L., \& {Wyart}, J.-F. 2000, \physscr, 61, 420

\bibitem[{{Clay} {et~al.}(2017){Clay}, {Burgess}, {Busemann},
  {Ruzi{\'e}-Hamilton}, {Joachim}, {Day}, \&
  {Ballentine}}]{2017Natur.551..614C}
{Clay}, P.~L., {Burgess}, R., {Busemann}, H., {et~al.} 2017, \nat, 551, 614

\bibitem[{{Cowley} \& {Wahlgren}(2006)}]{2006A&A...447..681C}
{Cowley}, C.~R. \& {Wahlgren}, G.~M. 2006, \aap, 447, 681

\bibitem[{{Curtis} {et~al.}(1984){Curtis}, {Martinson}, {Leavitt}, {Dietrich},
  {Bashkin}, \& {Denne}}]{1984PhLA..105..212C}
{Curtis}, L.~J., {Martinson}, I., {Leavitt}, J.~A., {et~al.} 1984, Physics
  Letters A, 105, 212

\bibitem[{{Dreizler} \& {Werner}(1996)}]{1996A&A...314..217D}
{Dreizler}, S. \& {Werner}, K. 1996, \aap, 314, 217

\bibitem[{{Gibbs} {et~al.}(1929){Gibbs}, {Vieweg}, \&
  {Gartlein}}]{1929PhRv...34..406G}
{Gibbs}, R.~C., {Vieweg}, A.~M., \& {Gartlein}, C.~W. 1929, Physical Review,
  34, 406

\bibitem[{{Hoyer} {et~al.}(2018){Hoyer}, {Rauch.}, {Werner}, \&
  {Kruk}}]{2018arXiv180102414H}
{Hoyer}, D., {Rauch.}, T., {Werner}, K., \& {Kruk}, J.~W. 2018, ArXiv e-prints

\bibitem[{{Joshi} \& {van Kleef}(1986)}]{1986PhyS...34..135J}
{Joshi}, Y.~N. \& {van Kleef}, T.~A.~M. 1986, \physscr, 34, 135

\bibitem[{{Kelly}(1987)}]{1987JPCRD..16S....K}
{Kelly}, R.~L. 1987, Journal of Physical and Chemical Reference Data, 16

\bibitem[{Kramida {et~al.}(2017)Kramida, {Yu.~Ralchenko}, Reader, \& {and NIST
  ASD Team}}]{NIST_ASD}
Kramida, A., {Yu.~Ralchenko}, Reader, J., \& {and NIST ASD Team}. 2017, {NIST
  Atomic Spectra Database (ver. 5.5.1), [Online]. Available:
  {\tt{https://physics.nist.gov/asd}} [2017, November 20]. National Institute
  of Standards and Technology, Gaithersburg, MD.}

\bibitem[{{Lang}(1927)}]{1927PNAS...13..341L}
{Lang}, R.~J. 1927, Proceedings of the National Academy of Science, 13, 341

\bibitem[{{Leckrone} {et~al.}(1999){Leckrone}, {Proffitt}, {Wahlgren},
  {Johansson}, \& {Brage}}]{1999AJ....117.1454L}
{Leckrone}, D.~S., {Proffitt}, C.~R., {Wahlgren}, G.~M., {Johansson}, S.~G., \&
  {Brage}, T. 1999, \aj, 117, 1454

\bibitem[{{Moore}(1971)}]{1971stas.book.....M}
{Moore}, C.~E. 1971, {Selected tables of atomic spectra - A: Atomic energy
  levels - Second edition - B: Multiplet tables; N IV, N V, N VI, N VII. Data
  derived from the analyses of optical spectra}

\bibitem[{{Morton}(2000)}]{2000ApJS..130..403M}
{Morton}, D.~C. 2000, \apjs, 130, 403

\bibitem[{{Napiwotzki} {et~al.}(1995){Napiwotzki}, {Hurwitz}, {Jordan},
  {Bowyer}, {Koester}, {Weidemann}, {Lampton}, \&
  {Edelstein}}]{1995A&A...300L...5N}
{Napiwotzki}, R., {Hurwitz}, M., {Jordan}, S., {et~al.} 1995, \aap, 300, L5

\bibitem[{{Rao} \& {Rao}(1934)}]{1934PPS....46..163R}
{Rao}, A.~S. \& {Rao}, K.~R. 1934, Proceedings of the Physical Society, 46, 163

\bibitem[{{Rauch} {et~al.}(2016){Rauch}, {Quinet}, {Hoyer}, {Werner},
  {Demleitner}, \& {Kruk}}]{rauchetal2016mo}
{Rauch}, T., {Quinet}, P., {Hoyer}, D., {et~al.} 2016, \aap, 587, A39

\bibitem[{{Rauch} {et~al.}(2017){Rauch}, {Quinet}, {Kn{\"o}rzer}, {Hoyer},
  {Werner}, {Kruk}, \& {Demleitner}}]{2017A&A...606A.105R}
{Rauch}, T., {Quinet}, P., {Kn{\"o}rzer}, M., {et~al.} 2017, \aap, 606, A105

\bibitem[{{Rauch} {et~al.}(2012){Rauch}, {Werner}, {Bi{\'e}mont}, {Quinet}, \&
  {Kruk}}]{test}
{Rauch}, T., {Werner}, K., {Bi{\'e}mont}, {\'E}., {Quinet}, P., \& {Kruk},
  J.~W. 2012, \aap, 546, A55

\bibitem[{{Riyaz} {et~al.}(2012){Riyaz}, {Tauheed}, \&
  {Rahimullah}}]{2012JQSRT.113.2072R}
{Riyaz}, A., {Tauheed}, A., \& {Rahimullah}, K. 2012, \jqsrt, 113, 2072

\bibitem[{{Riyaz} {et~al.}(2014){Riyaz}, {Tauheed}, \&
  {Rahimullah}}]{2014JQSRT.147...86R}
{Riyaz}, A., {Tauheed}, A., \& {Rahimullah}, K. 2014, \jqsrt, 147, 86

\bibitem[{{Wahlgren} \& {Leckrone}(2008)}]{2008CoSka..38..463W}
{Wahlgren}, G.~M. \& {Leckrone}, D.~S. 2008, Contributions of the Astronomical
  Observatory Skalnate Pleso, 38, 463

\bibitem[{{Warner} \& {Kirkpatrick}(1969)}]{1969MNRAS.142..265W}
{Warner}, B. \& {Kirkpatrick}, R.~C. 1969, \mnras, 142, 265

\bibitem[{{Werner} {et~al.}(2003){Werner}, {Deetjen}, {Dreizler}, {Nagel},
  {Rauch}, \& {Schuh}}]{2003ASPC..288...31W}
{Werner}, K., {Deetjen}, J.~L., {Dreizler}, S., {et~al.} 2003, in Astronomical
  Society of the Pacific Conference Series, Vol. 288, Stellar Atmosphere
  Modeling, ed. I.~{Hubeny}, D.~{Mihalas}, \& K.~{Werner}, 31

\bibitem[{{Werner} \& {Dreizler}(1999)}]{1999JCoAM.109...65W}
{Werner}, K. \& {Dreizler}, S. 1999, Journal of Computational and Applied
  Mathematics, 109, 65

\bibitem[{{Werner} {et~al.}(2012{\natexlab{a}}){Werner}, {Dreizler}, \&
  {Rauch}}]{tmap2012}
{Werner}, K., {Dreizler}, S., \& {Rauch}, T. 2012{\natexlab{a}}, {TMAP:
  T{\"u}bingen NLTE Model-Atmosphere Package}, Astrophysics Source Code Library
  [record ascl:1212.015]

\bibitem[{{Werner} {et~al.}(2012{\natexlab{b}}){Werner}, {Rauch}, {Ringat}, \&
  {Kruk}}]{2012ApJ...753L...7W}
{Werner}, K., {Rauch}, T., {Ringat}, E., \& {Kruk}, J.~W. 2012{\natexlab{b}},
  \apjl, 753, L7

\end{thebibliography}

\end{document}